\documentclass[fleqn,usenatbib]{mnras}

\usepackage{newtxtext,newtxmath,cuted,lipsum}
\usepackage[T1]{fontenc} 
\usepackage{graphicx}	
\usepackage{amsmath}	
\usepackage{amsfonts}
\usepackage{enumerate} 
\usepackage{colordvi} 
\usepackage{bm}
\usepackage{epsfig}
\usepackage{float}
\usepackage{verbatim}
\usepackage[dvipsnames]{xcolor}
\usepackage{multirow,multicol}
\usepackage{ulem}
\usepackage{hyperref}
\usepackage{fontawesome5} 
\usepackage{cleveref}

\newcommand{\github}[1]{%
   \href{#1}{\faGithub}%
}

\newcommand{\bfx}{\mbox{\boldmath$x$}}

\newcommand{\bfp}{\mbox{\boldmath$p$}}

\newcommand{\bfr}{\mbox{\boldmath$r$}}
\newcommand{\bfJ}{\mbox{\boldmath$J$}}
\newcommand{\bfF}{\mbox{\boldmath$F$}}
\newcommand{\bfu}{\mbox{\boldmath$u$}}

\newcommand{\Hh}{\mathcal{H}}
\newcommand{\n}{\nabla}

\newcommand{\wDE}{w}

\newcommand{\pN}{\partial_{\scriptscriptstyle N}}

\newcommand{\be}{\begin{equation}}
\newcommand{\ee}{\end{equation}}
\newcommand{\bea}{\begin{eqnarray}}
\newcommand{\eea}{\end{eqnarray}}

\title[Nonlinear spectrum for interacting dark energy]{On the road to percent accuracy VI: the nonlinear power spectrum for interacting dark energy with baryonic feedback and massive neutrinos}  

\author[P. Carrilho et al.]{Pedro Carrilho$^{1,2}$\thanks{E-mail: pedro.carrilho@ed.ac.uk},
Karim Carrion$^{3}$,
Benjamin Bose$^{4}$, 
Alkistis Pourtsidou$^{1,5,6}$,
Juan Carlos Hidalgo$^3$,
\newauthor
Lucas Lombriser$^4$,
Marco Baldi$^{7,8,9}$
\\
$^{1}$ Institute for Astronomy, The University of Edinburgh, Royal Observatory, Edinburgh EH9 3HJ, UK \\
$^{2}$ School of Physical and Chemical Sciences, Queen Mary University of London, Mile End Road, London E1 4NS, U.K. \\
$^3$ Instituto de Ciencias F\'{i}sicas, Universidad Nacional Aut\'{o}noma de M\'{e}xico,
Cuernavaca, Morelos, 62210, Mexico\\
$^4$ D\'epartement de Physique Th\'eorique, Universit\'e de Gen\`eve, 24 quai Ernest Ansermet, 1211 Gen\`eve 4, Switzerland. \\
$^{5}$ Higgs Centre for Theoretical Physics, School of Physics and Astronomy,
The University of Edinburgh, Edinburgh EH9 3FD, UK \\
$^{6}$ Department of Physics \& Astronomy, University of the Western Cape, Cape Town 7535, South Africa. \\
$^{7}$ Dipartimento di Fisica e Astronomia "Augusto Righi", Alma Mater Studiorum Universit\`{a} di Bologna, via Gobetti 93/2, I-40129 Bologna, Italy.\\
$^{8}$ INAF - Osservatorio di Astrofisica e Scienza dello Spazio di Bologna, via Gobetti 93/3, I-40129 Bologna, Italy. \\
$^{9}$ INFN - Sezione di Bologna, viale Berti Pichat 6/2, I-40127 Bologna, Italy. 
}
\date{Accepted XXX. Received YYY; in original form ZZZ}

\pubyear{2021}

\begin{document}
\label{firstpage}
\pagerange{\pageref{firstpage}--\pageref{lastpage}}
\maketitle

\begin{abstract}
Understanding nonlinear structure formation is crucial for fully exploring the data generated by stage IV surveys, requiring accurate modelling of the power spectrum. This is challenging for deviations from $\Lambda$CDM, but we must ensure that alternatives are well tested, to avoid false detections. We present an extension of the halo model reaction framework for interacting dark energy. We modify the halo model including the additional force present in the Dark Scattering model and implement it into {\tt ReACT}. The reaction is combined with a pseudo spectrum from {\tt EuclidEmulator2} and compared to N-body simulations. Using standard mass function and concentration-mass relation, we find predictions to be 1\% accurate at $z=0$ up to $k=0.8~h/{\rm Mpc}$ for the largest interaction strength tested ($\xi=50$ b/GeV), improving to $2~h/{\rm Mpc}$ at $z=1$. For smaller interaction strength ($10$ b/GeV), we find 1\% agreement at $z=1$ up to scales above $3.5~h/{\rm Mpc}$, being close to $1~h/{\rm Mpc}$ at $z=0$. Finally, we improve our predictions with the inclusion of baryonic feedback and massive neutrinos and study degeneracies between the effects of these contributions and those of the interaction. Limiting the scales to where our modelling is 1\% accurate, we find a degeneracy between the interaction and feedback, but not with massive neutrinos. We expect the degeneracy with feedback to be resolvable by including smaller scales. This work represents the first analytical tool for calculating the nonlinear spectrum for interacting dark energy models.
\end{abstract}


\section{Introduction}

Forthcoming large-scale structure (LSS) Stage IV surveys such as ESA's  \textit{Euclid} satellite mission\footnote{\url{www.euclid-ec.org}} \citep{Laureijs:2011gra, Blanchard:2019oqi} and  the Vera C. Rubin Observatory’s Legacy Survey of Space and Time (VRO/LSST)\footnote{\url{https://www.lsst.org/}} \citep{2018arXiv180901669T},
are aiming to probe the nature of dark energy by performing high-precision galaxy clustering and weak gravitational lensing measurements.

The standard cosmological model, $\Lambda$CDM, is currently providing the best fit to data from Cosmic Microwave Background (CMB) and LSS surveys (e.g. \citet{Aghanim:2018eyx, Anderson_2012, Song:2015oza, Beutler_2016, Troster:2019ean, Alam:2020sor, Abbott:2021bzy, Heymans_2021}). $\Lambda$CDM assumes that dark matter is cold (CDM), dark energy is a cosmological constant, $\Lambda$, and that General Relativity describes gravity on all scales. Therefore, the model requires a dark sector whose nature is unknown. Beyond this, there are now hints of tensions between various data sets \citep{Verde:2019ivm,Knox:2019rjx}, which may indicate new dark sector physics is required, beyond what is assumed within $\Lambda$CDM.

A particular discrepancy has been found between measurements of the growth of structure as estimated from early Universe measurements~\citep{Aghanim:2018eyx} and from late-time observations~\citep{Abbott:2020knk,Heymans_2021,Vikhlinin:2008ym,Reid:2012sw,Gil-Marin:2016wya,deHaan:2016qvy,Blake:2011rj,Beutler:2013yhm,Simpson:2015yfa,2021arXiv210513544S}. This relates to the amplitude of the late-time matter power spectrum, parameterised by $\sigma_8$, which is consistently measured by weak lensing surveys to be smaller than the one derived from CMB observations assuming $\Lambda$CDM. The tension could be a result of unaccounted systematic effects, but if this possibility is excluded then the tension points to new physics that is likely to require modifications to the dark sector. 

A number of alternatives to $\Lambda$CDM have been hypothesised, either replacing General Relativity (GR) with modified gravity (MG) on cosmological scales \citep[for reviews see][]{Clifton:2011jh,Joyce:2016vqv,Koyama:2018som} or generalizing the dark sector to include dynamical dark energy \citep[for reviews see][]{Copeland:2006wr,Li:2011sd}. Another interesting class of models to test are coupled models of dark matter and dark energy. These are traditionally called interacting dark energy (IDE) models and come in different flavours, depending on how energy is transferred between the dark sector species (see e.g. \citet{Amendola:1999er,Farrar:2003uw}). Most IDE models assume some form of energy transfer that affects the background and perturbations. These models usually fail to fit CMB data or are severely constrained~\citep{Bean_2008,Xia:2009zzb,Amendola_2012,Gomez-Valent:2020mqn} \citep[see, however, ][for more recent models evading these constraints]{Barros:2018efl,vandeBruck:2019vzd}. An alternative class of models assumes instead that the interaction exchanges only momentum at the level of the perturbations, allowing for very good fits to CMB and LSS data \citep{Simpson:2010vh, Pourtsidou:2013nha, Baldi:2016zom, Pourtsidou:2016ico, Mancini:2021lec}.
This feature could also be used to alleviate the $\sigma_8$ tension~\citep{Lesgourgues:2015wza,Linton:2017ged,Buen-Abad:2017gxg,Kase:2019mox,DiValentino:2019ffd, Amendola:2020ldb,2021arXiv210611222B}.

Future surveys will have unprecedented statistical power and for that reason, they have great potential to constrain or detect new physics. However, this will only be possible if we can improve our theoretical modelling of the nonlinear scales for deviations to the standard model to a sufficient precision.  This is already a difficult problem within $\Lambda$CDM, for which several standard physical effects complicate its predictive power at small scales. In addition to nonlinear gravitational collapse, this includes the effects of a non-zero neutrino mass \citep{2018MNRAS.481.1486B,2012MNRAS.420.2551B,2014JCAP...11..039B,2016MNRAS.459.1468M,2017ApJ...847...50L,Tram:2018znz,Massara:2014kba,2020arXiv200406245A}, as well as baryonic feedback processes (e.g., \citealt{vanDaalen2011,Mummery2017,Springel2018,vanDaalen:2019pst, 2021arXiv210904458T}; for a review see \citealt{Chisari:2019tus}). In particular, for Stage IV surveys, ignoring such effects can lead to biased estimates of cosmological parameters \citep[for examples with baryonic feedback see][]{Semboloni:2011fe,Schneider:2019snl, Martinelli:2020yto}. For beyond $\Lambda$CDM cosmologies, this problem is typically aggravated by the additional nonlinear effects of the modified cosmology, which need to be modelled to sufficient precision and may be degenerate with other effects. This illustrates the need for testing and validation of the nonlinear modelling developed for non-standard cosmologies. In the context of IDE, efforts in this direction have been made for spectroscopic galaxy clustering by \citet{Carrilho:2021hly}, where perturbative models were tested and it was shown that unbiased constraints are possible when appropriate scale cuts are chosen. Comparable modelling and testing is currently missing for weak lensing observables in the context of IDE, which is what motivates this work.

A very promising and flexible method for modelling nonlinearities for non-standard cosmologies is the {\it halo model reaction} formalism \citep{Cataneo:2018cic}. This novel method has been able to reach percent level accuracy at predicting the power spectrum for various models beyond $\Lambda$CDM. This led to the development of {\tt ReACT} \citep{Bose:2020wch}, a code that can efficiently predict the halo model reaction and therefore be used in data analyses. {\tt ReACT} has already been utilised in the KiDS-1000 weak lensing data analysis for constraining modified gravity \citep{Troester:2020}. In addition, in \cite{Cataneo:2019fjp, 2021MNRAS.508.2479B}, the reaction formalism was developed for massive neutrinos and applied to {\tt ReACT}, which allowed for percent-level predictions of the nonlinear power spectrum in scenarios with both modified gravity and massive neutrinos. Furthermore, baryonic effects were also successfully included in \cite{2021MNRAS.508.2479B}, using parameterised feedback models, such as {\tt HMCODE} \citep{Mead:2020vgs} and were tested against hydrodynamical simulations. This was the first formalism that can accurately include the nonlinear effects of massive neutrinos for beyond-$\Lambda$CDM cosmologies, as well as strongly suggesting that baryonic feedback can be reliably modelled independently from  massive neutrinos and dark energy.

In this paper, we present an extension to the framework of \cite{Cataneo:2018cic} to include the effects of interactions in the dark sector. We modify the reaction formalism to include the additional force generated by the interaction, taking into account both the effects on gravitational collapse and virialisation of structures. This represents a new tool to calculate predictions for interacting models, which will permit them to be tested by future high-precision surveys.

This paper is organised as follows: In \autoref{sec:ide} we review the theory of IDE, establish the different classes of models and specify the particular models that we focus on; \autoref{sec:reaction} is dedicated to the reaction formalism and the modifications introduced in order to model IDE; in \autoref{sec:results} we show the comparisons of our modelling with simulations and demonstrate its accuracy; and in \autoref{sec:bar_neu} we extend our modelling to include baryonic feedback and massive neutrinos and explore degeneracies between their nonlinear effects and those of IDE. Finally, we summarize our conclusions in \autoref{sec:summary}.


\section{Interacting Dark Energy}\label{sec:ide}

\subsection{General interacting dark energy modelling}

Given that the nature of the two dark sectors
is still unknown, it may be that dark matter and
dark energy can be coupled to each other (see, for instance, \citet{Amendola:1999er,Pourtsidou:2013nha, Tamanini:2015iia, DiValentino:2019jae}). Traditionally interacting dark energy (IDE) has been described as a scalar field ($\phi$) explicitly coupled to cold dark matter (c). We can then define a coupling current $J_\mu$ and the Bianchi identities take the form:
\bea
    \nabla_\nu T^\nu_{({\rm c})\mu} = J_\mu 
    = - \nabla_\nu T^\nu_{(\phi)\mu} \, ,
\eea
so that the total energy-momentum tensor of the dark components is conserved.
It is also assumed that baryons remain uncoupled, which evades the need for screening mechanisms at solar-system scales.

As the form of the coupling current $J_\mu$ is a phenomenological choice, a plethora of different models can be constructed. In \citet{Pourtsidou:2013nha}, three distinct types of models of scalar field dark energy coupled to dark matter were constructed using a Lagrangian approach and the pull-back formalism for fluids. In a subsequent paper \citep{Skordis:2015yra}, the authors developed the Parameterised Post-Friedmannian (PPF) framework for IDE. Both methods showed that the most popular coupled quintessence models (e.g. \citet{Amendola:1999er}) belong to a small subset of more general classes of theories given by the action \citep{Pourtsidou:2013nha}
\bea
    S=\int{\rm d}^4 x \, \sqrt{-g}\, \mathcal{L}(n,\phi,X, Z)\,,
\eea 
in which $n$ is the dark matter number density, representing the degrees of freedom of the dark matter fluid, $X=\tfrac{1}{2}(\partial\phi)^2$ and $Z=u^\mu \partial_\mu \phi$, with $u^\mu$ the 4-velocity of the dark matter fluid. Coupled quintessence is a model of type 1, having no dependence on $Z$ and a Lagrangian of the form $\mathcal{L}=F(X,\phi)+f(n,\phi)$. Type 2 models introduce $Z$ coupled to $n$ via $\mathcal{L}=F(X,\phi)+f(n,Z)$, whereas in the type 3 case, $Z$ couples instead to the scalar field, with the Lagrangian $\mathcal{L}=F(X,\phi,Z)+f(n)$. 

Type 3 models are particularly interesting because the coupling only involves the dark matter velocity and not its density, resulting in $J^\mu\perp u^\mu$ and therefore allowing for pure momentum exchange in the dark sector. The main consequence of this is that there is no coupling at the background level ($\bar J^\mu=0$), with only the fluctuations being affected by the interaction. In addition, the dark matter energy-conservation equation remains uncoupled even at the level of the perturbations. Therefore, the model provides for a pure momentum-transfer coupling for perturbations. This is in contrast to the most common coupled dark energy, but it is also what makes this type of model able to fit CMB and LSS data very well \citep{Pourtsidou:2016ico, Baldi:2014, Baldi:2016zom,  Chamings:2019kcl,Amendola:2020ldb,Kase:2020hst, Mancini:2021lec}. 

An example sub-case of these pure momentum transfer theories is one where the Lagrangian for the scalar field $\phi$ is of the form $F= X+V(\phi)-\beta Z^2$, such that the action in an FLRW background can be written as:
\bea
 \nonumber 
S_{\phi}&=&
\int{\rm d}t \, {\rm d}^3x \, a^3 \left[
    \frac{1}{2}(1 - 2 \beta)  \dot{\phi}^2 - \frac{1}{2} |\nabla\phi|^2 - V(\phi)
\right] \, ,
\label{eq:action}
\eea
where a dot denotes a derivative with respect to cosmic time and $a$ is the scale factor. The model is physically acceptable for $\beta < 1/2$. For $\beta \rightarrow 1/2$ there is a strong coupling pathology, while for $\beta>1/2$ there is a ghost in the theory since the kinetic term becomes negative. In \citet{Pourtsidou:2016ico, Mancini:2021lec} it was shown that this model can fit CMB and LSS data for a wide range of the coupling parameter $\beta$.

This class of theories can be connected to the so-called {\it Dark Scattering} model, first constructed in \citet{Simpson:2010vh}, which also exhibits a pure momentum coupling. While it cannot be exactly mapped to a specific type 3 model, it was shown in \citet{Baldi:2016zom} that this can be done approximately, following the PPF formalism of \citet{Skordis:2015yra}. As we show in the next section, this model can be thought of as an extension of $w$CDM, with a well defined $\Lambda$CDM limit, thus being an attractive non-standard model to be tested with Stage-IV galaxy clustering and weak lensing surveys.

\subsection{The Dark Scattering model}

We now focus on the Dark Scattering model first developed in \citet{Simpson:2010vh}. This model is based on the assumption that the interaction between dark matter and dark energy is due to the scattering of their constituent particles. This is a phenomenological model, which shares features with dark radiation interaction models \citep{Lesgourgues:2015wza}, as well as with the type 3 momentum-exchange models described above \citep{Pourtsidou:2013nha, Skordis:2015yra}.
This model is built in analogy to Thomson scattering between charged particles and photons. This is evident when we identify the interaction terms as
\bea
& \bfJ_{T}= -\frac{4}{3}  \sigma_{\rm T}  a \rho_\gamma  n_e (\bfu_{e}-\bfu_{\gamma} ) \,, \\
& \downarrow \nonumber \\
& \bfJ_{{\rm DS}}=-(1+\wDE) \sigma_{\rm DS} a \rho_{\rm DE} n_{\rm c}  (\bfu_{{\rm c}}-\bfu_{{\rm DE}} ) \, ,\label{J_DS}
\eea
where symbols in bold represent 3-vectors, $\bfu_{X}$ denotes the peculiar velocity of species $X$; both cold dark matter (c) and dark energy ($\rm DE$) are assumed to be fluids, with the former having number density denoted by $n_{\rm c}$ and the latter having density $\rho_{\rm DE}$ and pressure $P_{\rm DE}=\wDE\rho_{\rm DE}$, thus defining the equation of state parameter $\wDE$. We denote the dark scattering cross section as $\sigma_{\rm DS}$, which is assumed constant. We work in the Newtonian approximation, which takes velocities to be non-relativistic and gravitational fields to be weak. Under this approximation the interaction in \autoref{J_DS} receives no nonlinear corrections.

With this interaction, the linear Euler equations for the interacting dark components are
\bea
    \theta_{\rm c}'+\Hh\theta_{\rm c}+\n^2\Phi=(1+\wDE) \frac{\rho_{\rm DE}}{\rho_{\rm c}}an_{\rm c}\sigma_{\rm DS}(\theta_{\rm DE}-\theta_{\rm c})\,,\\
    \theta_{\rm DE}'-2\Hh\theta_{\rm DE}-\frac1{1+\wDE}\n^2\delta_{\rm DE}+\n^2\Phi=an_{\rm c}\sigma_{\rm DS}(\theta_{\rm c}-\theta_{\rm DE})\,,
\eea
where a prime denotes a derivative with respect to conformal time, $\Hh=a'/a$ is the conformal Hubble rate, $\theta_X \equiv \n\cdot \bfu_X$ are the divergences of the velocities, $\Phi$ is the gravitational potential and $\delta_X = \delta\rho_X/\rho_X$ is the density contrast of species $X$. We assume the sound speed of dark energy fluctuations is $c_s^2=1$. This implies that on sub-horizon scales, the dark energy fluctuations are heavily damped, such that they can be neglected for the evolution of dark matter, resulting in a simplified Euler equation:
\bea
    \theta_{\rm c}'+\Hh(1+A)\theta_{\rm c}+\n^2\Phi=0\,,
    \label{Euler_simple}
\eea
where we have defined the variable $A$ as
\bea
A \equiv   \xi \left(1+\wDE\right) \dfrac{3\Omega_{\rm DE}}{8\pi G} H \, ,
\label{Eq:interaction}
\eea
Here we have introduced the coupling parameter $\xi \equiv \sigma_{\rm DS}/m_{\rm c} $ in units $[\rm {b}/\rm GeV]$, which encodes information on the CDM particle mass $m_{\rm c}$ and the scattering cross section $\sigma_{\rm DS}$, thereby modulating the strength of the interaction. We also use here the Hubble rate $H=\dot a/a$. \autoref{Euler_simple} reveals that the only effect of the interaction is to introduce additional friction in the evolution of the velocities. There are no other effects relative to a $w$CDM model, as the energy-conservation equation remains unchanged. In addition, there is a well defined $\Lambda$CDM limit, which is the same as for $w$CDM: when $\wDE\rightarrow -1$, we have $A=0$.

In the context of the PPF formalism developed by \citet{Skordis:2015yra}, this model can be shown not to be exactly of type 3, since in that case one has 
\bea
\bfJ=B_3 \n\delta_{\rm DE}+B_5 \bfu_{{\rm DE}}+ B_6 \bfu_{{\rm c}}\,,
\eea
with all $B_I$ being different and only vanishing if there is no interaction. However, \citet{Baldi:2016zom} showed that this can be done approximately, using a Lagrangian of the type $F\propto \exp(-Z)$. Alternatively, if the sound speed is $c_s^2=1$ then the contributions from dark energy fluctuations can be neglected, making those type 3 models have a similar form to the Dark Scattering case, with equivalence being achieved if the time-dependence of $B_6$ can be matched to the one from 
\autoref{J_DS}.

Using N-body simulations, the Dark Scattering model has been extended to the nonlinear level in \citet{Baldi:2014}, by assuming that the interaction is always linear in the dark matter velocity, therefore allowing one to extrapolate its effect for the equation of motion of individual particles as
\bea
\bfx'' = - \Hh (1+A) \bfx' - \nabla \Phi \, ,
\label{Eq:interaction2}
\eea
where $\bfx'$ is the velocity of particles in comoving coordinates. As mentioned above, this assumption of linearity is justified under the Newtonian approximation.\footnote{It is likely this approximation of linearity in the dark matter velocity also holds for many of the type 3 theories that are sufficiently similar to Dark Scattering. For those theories, the nonlinear modelling developed in this work is expected to also apply when written in terms of the generic coupling function $A$.} This formulation of the theory was used to simulate the nonlinear evolution of dark matter in \citet{Baldi:2014, Baldi:2016zom} and used to evaluate the matter power spectrum for different values of the coupling parameter $\xi$. We will also use this formulation to develop the halo model reaction formalism for this theory in the next section, which we will compare to those simulations.

\section{The halo model reaction}\label{sec:reaction}

We briefly review the {\it halo model reaction} approach to modelling the nonlinear matter power spectrum for general cosmologies \citep{Cataneo:2018cic,Cataneo:2019fjp}. In this scheme the nonlinear spectrum is given as a product of two key quantities, the halo model reaction  $\mathcal{R}(k,z)$ and the nonlinear {\it pseudo} power spectrum $P^{\rm pseudo}_{\rm NL} (k,z)$

\bea
    P_{\rm NL}(k,z) = \mathcal{R}(k,z)  \times  P^{\rm pseudo}_{\rm NL} (k,z) \, . \label{eq:nlps}
\eea
The pseudo cosmology is defined as a $\Lambda$CDM cosmology whose initial conditions are tuned such that the linear clustering at the target redshift matches the target cosmology. In this way, the reaction models the {\it nonlinear modifications} to the power spectrum arising from beyond $\Lambda$CDM physics, be it modifications to general relativity, massive neutrinos or a non-standard dark sector. 

This quantity is given by \citep{2021MNRAS.508.2479B}
\bea
    \mathcal{R}(k)=\frac{\bar{f}_{\nu}^{2} P_{\mathrm{HM}}^{(\mathrm{cb})}(k)+2 f_{\nu}\bar{f}_{\nu} P_{\mathrm{HM}}^{(\mathrm{cb} \nu)}(k)+f_{\nu}^{2} P_{\mathrm{L}}^{(\nu)}(k)}{P_{\mathrm{L}}^{(\mathrm{m})}(k)+P_{\mathrm{1h}}^{\mathrm{pseudo}}(k)}  \, ,
    \label{eq:reaction}
\eea
where we have dropped the redshift dependence for brevity. The superscript $\rm (m) \equiv (cb+\nu) $ with `cb' standing for CDM plus baryons and `$\nu$' standing for massive neutrinos. The combination (cb$\nu$) is approximated by $P_{\mathrm{HM}}^{(\mathrm{cb} \nu)}(k) \approx \sqrt{P_{\mathrm{L}}^{(\nu)}(k) P_{\mathrm{HM}}^{(\mathrm{cb})}(k)}$. 
 The subscript `HM' stands for halo model and $\bar{f}_{\nu} \equiv (1-f_\nu)$ with  $f_\nu = \Omega_{\nu ,0}/ \Omega_{\rm m,0}$ being the ratio of present day massive neutrino density to the total matter density. The subscripts `L' stand for linear and `1h'  stands for 1-halo. We refer the reader to the review by \cite{Cooray:2002dia} for details on how the 1-halo spectrum is constructed. The linear spectra, which include the beyond $\Lambda$CDM physics, can all be produced using a Boltzmann solver such as {\tt CAMB} \citep{Lewis:2002ah} or {\tt CLASS} \citep{2011arXiv1104.2932L} and the extensions thereof (e.g, \cite{Zucca:2019xhg,2011JCAP...08..005H,2009PhRvD..79h3513Z}). In particular, we have modified {\tt CLASS} for the Dark Scattering model and we use it for many of our calculations. 

From \autoref{eq:reaction} we see that this nonlinear correction is simply a ratio of the target-to-pseudo halo model power spectra with the effects of massive neutrinos added in linearly in the numerator, which was shown to be a sufficient approximation in \cite{Castorina:2015bma}. One may now ask, why correct a nonlinear pseudo spectrum and not a full $\Lambda$CDM spectrum? The importance of using the pseudo cosmology comes from the observation that the mass function in both target and pseudo cosmologies becomes far more similar as they share the same linear clustering that enters the peak statistic \citep{Mead:2016ybv}. This similarity provides a smoother transition between the 2-halo and 1-halo regimes, an issue of previous halo model prescriptions \citep{Cooray:2002dia,Cacciato:2008hm,Giocoli:2010dm}. 

Finally, the full halo model reaction as described in \cite{Cataneo:2018cic} and \cite{2021MNRAS.508.2479B} also involves a perturbation theory-based calculation, entering in the $P_{\mathrm{HM}}^{(\mathrm{cb})}(k)$
term. This term is only relevant when new mode couplings are introduced in the modification to $\Lambda$CDM, typical of theories of gravity beyond general relativity (see \cite{Koyama:2018som} for a review on modified gravity and screening mechanisms). Since we will only be considering a modified dark sector here, this additional term will not be used. 

\subsection{Nonlinearities for interacting dark energy}
\subsubsection{Evolution equations} 
The late-time evolution of the Universe is assumed to be well described by a perturbed flat Friedmann-Lema\^{i}tre-Robertson-Walker (FLRW) spacetime, whose contents are baryons, neutrinos, CDM and DE. At the background level, there is no modification with respect to the $w$CDM model, such that the energy conservation equations for each species $X$ can be written in general form as
\bea
\dot\rho_X=-3 H (1+w_X) \rho_X\,.
\eea
The Friedmann equation can then be expressed as,
\bea
H^2 = H^2_0\left(\Omega_{{\rm m},0} a^{-3} + \Omega_{{\rm DE},0} e^{-3\int(1+\wDE){\rm d} \log a} \right)\, ,
\label{Eq:H_background}
\eea
where we have collected all non-relativistic species into $\Omega_{\rm m}$ and substituted the solutions for the energy densities. We use the Chevalier-Polarski-Linder (CPL) parametrisation \citep{Chevallier:2000qy,Linder:2002et} for the evolution of the dark energy equation of state, given by
\bea
\wDE=w_0+w_a(1-a)\,,
\label{param:cpl}
\eea
with $w_0$ and $w_a$ being constant parameters. In the case of $w_a=0$, we use $w$ as a parameter, for simplicity.

We treat the evolution of matter fluctuations in the Newtonian approximation, such that the nonlinear evolution equations for dark matter are\footnote{These can be obtained directly from \autoref{Eq:interaction2} using the suitable Vlasov equation for dissipative systems \citep[see][]{perepelkin2018vlasov},
\bea
  f'
+ \nabla_{\bfx}  \cdot \left( f \bfx' \right)
+ \nabla_{\bfp} \cdot \left( f  \bfp'\right) = 0 \, .
\label{Eq:Vlasov_classical2}
\eea}
\bea
&\delta_c' + \nabla \cdot \left((1+\delta_c) \bfu_c\right)=  0 \,, \label{Eq:continuity} \\
&\bfu_c'
+ (\bfu_c \cdot \nabla) \bfu_c
+ \mathcal{H} \bfu_c + \nabla \Phi = - A \mathcal{H} \bfu_c \, ,
\label{Eq:Euler}
\eea
where we have neglected pressure and other stress contributions that are typically small in the context of cold dark matter. Baryons and neutrinos have the same evolution equations as in $\Lambda$CDM, so we do not repeat them here. The system of equations is closed by the Poisson equation
\bea
& \nabla^2 \Phi(\bfx, \eta) = \frac{3}{2} \Omega_{\rm m} \mathcal{H}^2 \delta(\bfx, \eta) \, , 
\label{Eq:Poisson}
\eea
where $\delta = \delta_{\rm m}$ denotes the density contrast of all non-relativistic matter.

The Dark Scattering interaction is not universal, given that only dark matter interacts but other species do not. For this reason, one expects the effective strength of the interaction to be modulated by the dark matter fraction of the total matter, $f_c=\rho_c/\rho_{\rm m}$. This can be demonstrated by analysing the Euler equations of non-relativistic species at the linear level. We split those species into CDM and non-CDM, with the latter including baryons and massive neutrinos, which we approximate to evolve in the same way. We then define the total matter velocity divergence $\theta_{\rm m}\equiv f_c \theta_c+f_{b\nu}\theta_{b\nu}$ and the velocity difference $\Delta\theta\equiv\theta_c-\theta_{b\nu}$, whose evolution equations are
\bea
&\theta_{\rm m}'=-\Hh(1+A f_c)\theta_{\rm m}-\Hh A f_c f_{b\nu}\Delta\theta-\n^2\Phi\,,\\
&\Delta\theta'=-\Hh(1+A f_{b\nu})\Delta\theta-\Hh A\theta_{\rm m}\,.
\eea
Since $\Delta\theta$ is expected to be small, we can approximate the solution for the $\Delta\theta$ equation by its equilibrium solution (i.e. the solution which gives $\Delta\theta'=0$). Substituting that into the $\theta_{\rm m}$ equation results in
\bea
\theta_{\rm m}'=-\Hh\left(1+\frac{A f_c}{1+Af_{b\nu}}\right)\theta_{\rm m}-\n^2\Phi\,.
\eea
This demonstrates that the total matter evolves with an effective coupling function that depends on the relative amount of dark matter in the Universe. While this coupling function has a slightly modified time-dependence relative to the standard coupling function, $A$, we have verified with a numerical solution from our modified version of {\tt CLASS} that it is a very good approximation to evaluate the denominator at $z=0$, so that we can define an effective coupling constant $\bar\xi$, given by
\bea
\bar\xi=\frac{f_c}{1+A_0(1-f_c)}\xi\,.
\label{eff_xi}
\eea
This means that a single fluid description is possible by using this effective coupling to compute the growth factor and growth rate. At the nonlinear level, this approximation assumes that we can put together the clustering species when computing the halo model prediction by using this effective coupling. We use this assumption in the following sections, which describe the ingredients of the halo model that are modified when the interaction is active.
\subsubsection{Spherical Collapse}
In order to approximate the halo formation of the IDE model we use the Press-Schechter formalism \citep{Press:1973iz}, thus modelling it with spherical collapse. We follow the evolution of a spherical overdensity of physical radius $r$ and mass $M$, modelled through the well-known top-hat model \citep[see, e.g.][]{Schmidt2008},   
\bea
M = \dfrac{4 \pi}{3} r^3 \bar{\rho}_{\rm m} (1+\delta)  = \text{const.}
\label{Eq:mass}
\eea
For this idealised overdensity, the following nonlinear evolution is derived from \hyperref[Eq:continuity]{Equations~\ref*{Eq:continuity}} -- \ref{Eq:Poisson},
\bea
\ddot{\delta} + 2 H \dot{\delta} - \dfrac{4}{3}\dfrac{\dot{\delta}^2}{(1+\delta)} = \dfrac{(1+\delta)}{a^2} \nabla^2 \Phi - A H \dot{\delta} \, .
\label{Eq:delta_spherical_dot}
\eea
This can be rewritten in terms of the sphere radius $r$ instead of $\delta$, arriving at,
\bea
\dfrac{\ddot{r}}{r} + AH \dfrac{\dot{r}}{r} =  - \dfrac{4\pi G}{3} \left[\bar{\rho}_{\rm m} + (1+3w)\bar{\rho}_{\rm DE} \right] + AH^2 - \dfrac{4\pi G}{3} \bar{\rho}_{\rm m} \delta \, .
\label{Eq:r_ddot3}
\eea
We see that there are two contributions from the dark interaction, one which dampens the collapse and one that appears to enhance it. This split into two terms is due to using physical radius $r$ as a coordinate, instead of a comoving coordinate. This is explained by the fact that the interaction is only dependent on the peculiar velocity of dark matter, defined relative to the Hubble flow so the term on the right-hand-side of \autoref{Eq:r_ddot3} serves to compensate for the expansion. This can be clearly seen by using the comoving variable $y \equiv \frac{r}{r_i}  \frac{a_i}{a}$, which will also be useful for our numerical implementation, resulting in
\bea
\pN^2 y + \left(2 + A + \frac{\pN H}{H} \right)\pN y +  \left(\frac{H_0^2}{H^2} \frac{\Omega_{{\rm m},0}}{2a^3} \delta \right)\,  y = 0 \, ,
\eea
where $\pN$ denotes a derivative with respect to the number of e-folds, $N\equiv\ln a$. The density contrast of the top-hat is related to $y$ via 
\bea
\delta(y,a) = [1+\delta_i] \left(\frac{a_i}{a}y \right)^{-3} -1 \, ,
\label{y:delta_pert}
\eea
where $\delta_i$ is its initial value, when the physical size of the fluctuation is $r_i$ at the initial time $a_i$. While, according to the equations above, evolution would lead to full collapse to $r=0$, in the real Universe, a collapsing fluctuation will instead reach virial equilibrium. Therefore we evolve \autoref{Eq:r_ddot3} up to collapse to find the scale factor at that time, $a_{\rm col}$, while also determining at which point along the evolution virial equilibrium is reached, $a_{\rm vir}$, as well as the virial overdensity,
\bea
\Delta_{\rm vir} = [1+\delta(y,a_{\rm vir})]\left(\dfrac{a_{\rm col}}{a_{\rm vir}} \right)^3 \, ,
\label{y:virial_delta}
\eea 
and the corresponding mass
\bea
M_{\rm vir} = \dfrac{4 \pi}{3} r_{\rm vir}^3 \, \bar{\rho}_{\rm m}\, \Delta_{\rm vir} \, .
\label{y:virial_mass}
\eea
For that reason, we also need to know how the virial theorem is modified due to the dark sector interaction and we describe that in the next section.
\subsubsection{Virial Theorem}
As shown previously, the effect of the dark sector interaction is to generate an additional friction force on dark matter particles. Within the approximations used in this work, this friction force is clearly not conservative and it cannot be included in the traditional potential term of the virial theorem. Therefore, we must add a non-conservative force $\bfF^{\rm fric}$ to the standard expression
\bea
2\langle T\rangle+\langle W\rangle+\sum_i\langle \bfF^{\rm fric}_i\cdot \bfr_i\rangle=0 \, , 
\eea
where $T$ is the total kinetic energy of the system, $W$ is the potential term and the sum is over all particles, with physical positions $\bfr_i$.
The friction force that comes from the interacting term is 
\bea
\bfF^{\rm fric}_i= -m A H \bfx'_i \, .
\eea
We then integrate this with the distribution function over the momentum and position to obtain the total contribution, 
\bea
W_{\rm DS}=-A H\int {\rm d}^3x\,a\rho \bfu\cdot \bfx \,,
\eea
where the integral is over the comoving coordinate $\bfx$. Expressing this for the top hat profile described in the previous section, in terms of the normalised comoving radius $y$, we find the following expression for the contribution of the interaction to the virial theorem 
\bea
W_{\rm DS} = -\frac{3}{5} A  M (r_i H)^2 y \, \pN y \left(\dfrac{a}{a_i} \right)^2 \, .
\eea
Following \cite{Cataneo:2019fjp}, we write the contributions to the Virial theorem in units of $E_0 \equiv \frac{3}{10}M \left(H_0 r_i \right)^2$. The expressions in this case are: 
\bea
 & \dfrac{W_{\rm N}}{E_0} = - \Omega_{\rm m} \left( \dfrac{a^{-1}}{a^2_i}\right) y^2 (1+\delta) \, , \\
 & \dfrac{W_{\rm DE}}{E_0} = - \dfrac{H^2}{H^2_0}(1+3 w) \Omega_{\rm DE}\left(\dfrac{a}{a_i} \right)^2 y^2 \, ,\\
 & \dfrac{W_{\rm DS}}{E_0} = - 2 A \dfrac{H^2}{H^2_0} \left(\dfrac{a}{a_i} \right)^2  y \, \pN y \,  ,
\eea
where we have split the Newtonian contribution, $W_{\rm N}$ from that coming from the background acceleration, $W_{\rm DE}$. For completeness, we present also the total kinetic energy of the top hat, $T$,  
\bea
\dfrac{T}{E_0}=\dfrac{H^2}{H^2_0}\left[\dfrac{a}{a_i} (\pN y + y) \right]^2 \, .
\eea
To summarize, the equation that must be solved to find the virialisation time $a_{\rm vir}$, as well as the corresponding overdensity $\Delta_{\rm vir}$, is 
\bea
2T + W_{\rm N} + W_{\rm DE} + W_{\rm DS} = 0 \, .
\label{y:virial_theorem}
\eea
These modifications are then applied to the calculation of the halo model reaction using the code {\tt ReACT}. This results in predictions for the power spectrum for the dark scattering model that can be compared to simulations. We perform this validation in the next section.
\begin{figure*}
    \centering
    \includegraphics[width=1.0\textwidth]{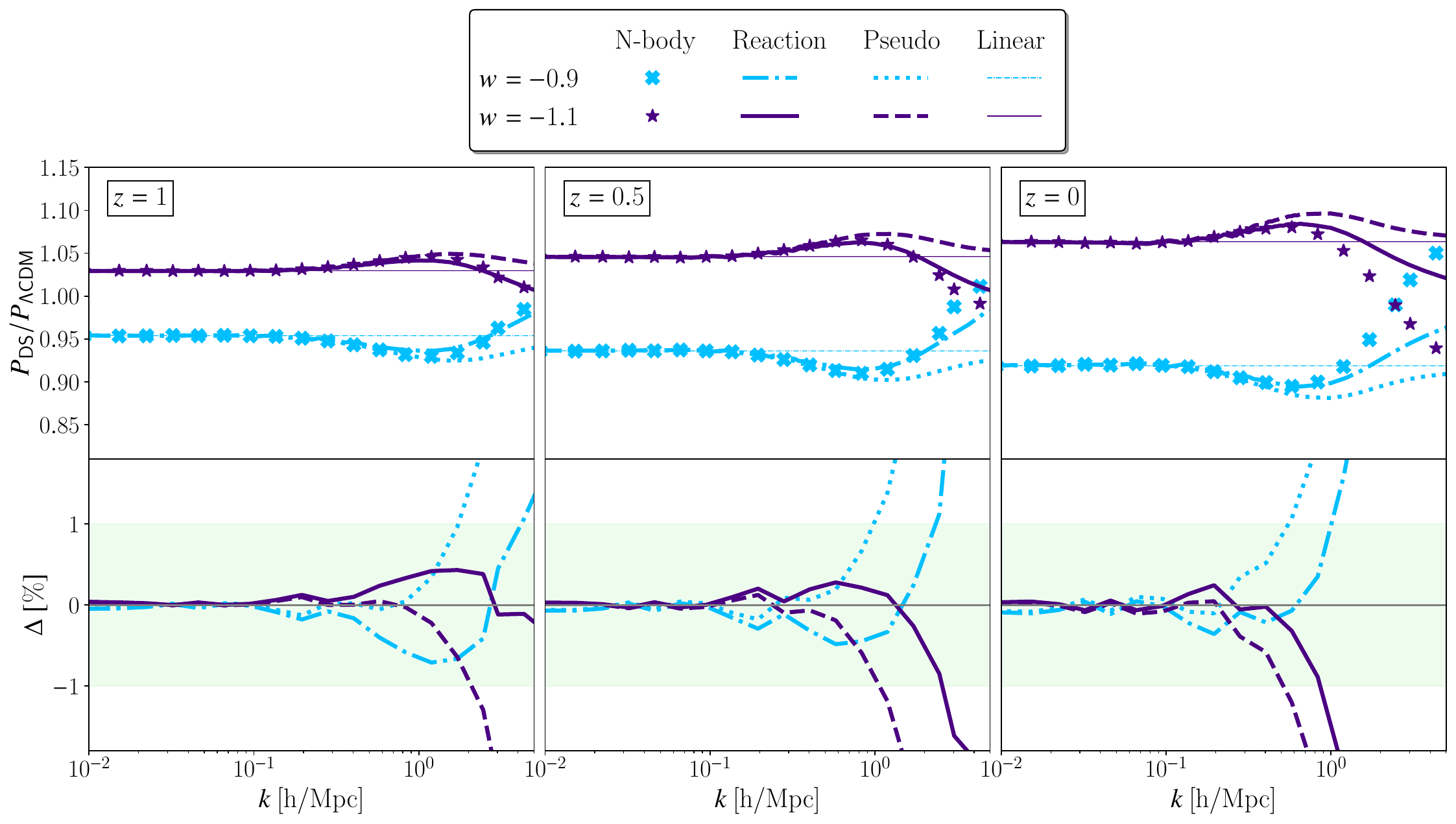}
    \caption{{\bf Top:} Ratios of Dark Scattering spectra to $\Lambda$CDM, for a coupling strength of $\xi = 10 \, \rm {b} /\rm GeV$. The blue curves are for $w = -0.9$, where crosses, dash-dotted lines, dotted lines and thin dashed-lines are measurements from simulations, the halo model reaction prediction, the pseudo spectrum prediction and the linear theory prediction, respectively. In purple, we show the results for $w = -1.1$ where stars, solid lines, dashed-lines and thin solid lines represent the same quantities for this case respectively. {\bf Bottom:} The residuals in percentage, $\Delta = 100\,\mathcal{\%} \cdot \left( 1 - R_{\rm prediction}/R_{{\rm N\text{-}body}}\right) $, for the reaction and pseudo spectrum predictions, where $R=P_{\rm DS}/P_{\Lambda{\rm CDM}}$ is the ratio shown in the top plot.}
    \label{fig:wCDM}
\end{figure*}
\begin{figure*}
    \centering
    \includegraphics[width=1.0\textwidth]{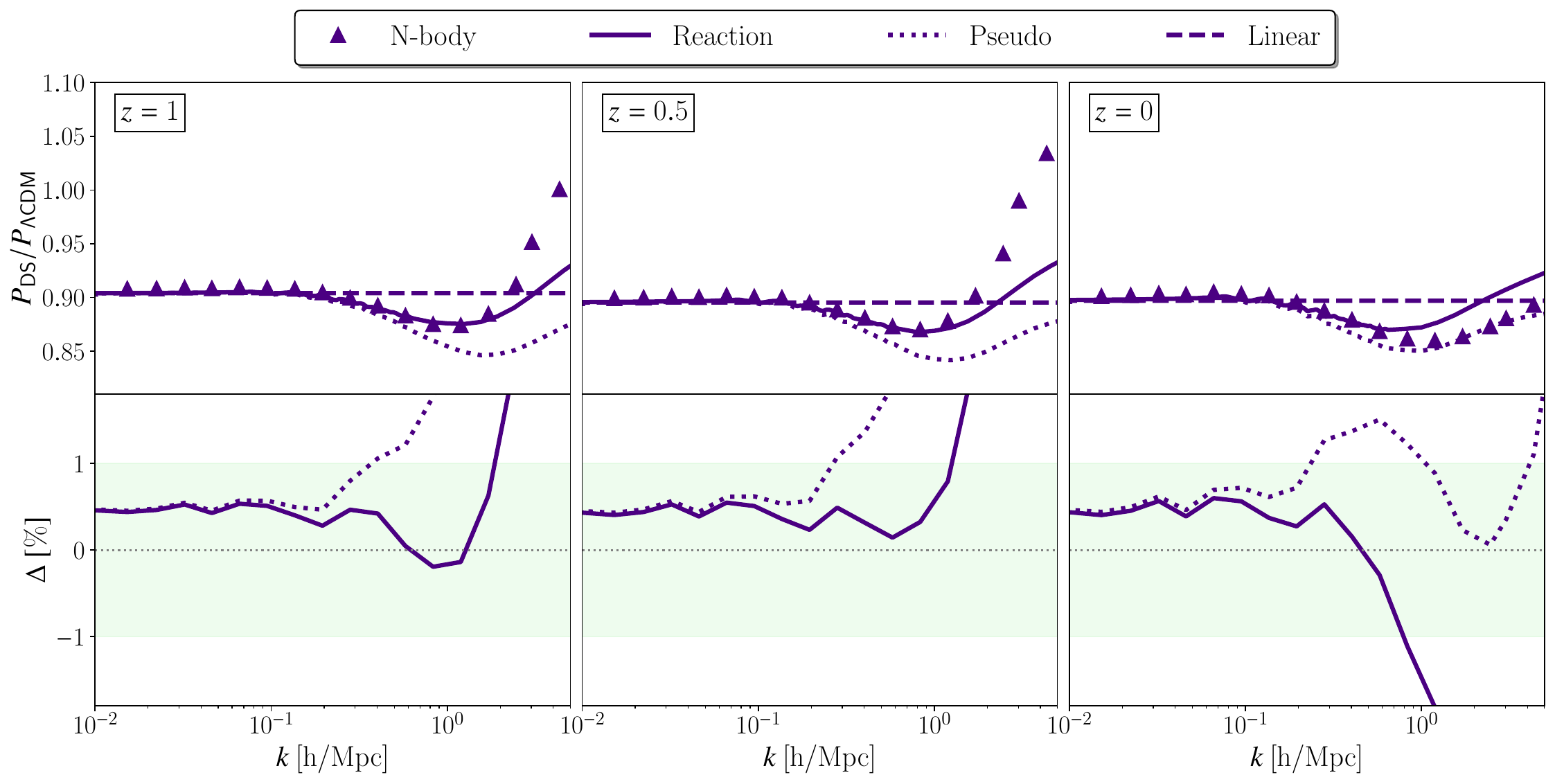}
    \caption{{\bf Top:} Ratio of Dark Scattering spectrum (CPL: $w_0=-1.1, w_a = 0.3$) with a value of $\xi = 50 \, \rm {b} /\rm GeV$ to $\Lambda$CDM. Points are measurements from the simulations whereas solid lines are the halo model reaction prediction. Dotted lines are the pseudo spectrum and dashed lines are the linear theory prediction. {\bf Bottom:} The residuals, as defined in \autoref{fig:wCDM}, for the reaction and pseudo spectrum predictions. Note that there is a 0.5\% discrepancy on large scales. The reason for this is that the N-body simulations have been found to capture linear growth with differing accuracy depending on the value of $\xi$ \citep{Baldi:2014}, so that the ratio between DS and $\Lambda$CDM is slightly modified with respect to the prediction. This is noticeable here and not in \autoref{fig:wCDM}, because of the much larger coupling used here.}
    \label{fig:CPL}
\end{figure*}
\section{Accuracy validation of the modelling against simulations}\label{sec:results} 

Here we will validate the halo model reaction predictions against the N-body simulations for the Dark Scattering model described in \cite{Baldi:2016zom}. These were performed using a modified version of the {\tt GADGET-2} N-body code \citep{Springel:2005mi} which consistently implements the momentum exchange between the dark matter particles and the underlying homogeneous dark energy field. These simulations contain $1024^3$ dark matter particles in a box of $1$ Gpc$/h$ per side. They begin the particle evolution at $z_{i}=99$ and trace it up to $z=0$. The resulting CDM particle mass is  $m_{c} = 8\times 10^{10}$ M$_{\odot }/h$ and the spatial resolution is $\epsilon = 24$ kpc$/h$ (equivalent to $k_{\epsilon} = 261~h/\mathrm{Mpc}$).

In this work we consider a subset of the simulations presented in \cite{Baldi:2016zom}, which we summarise in \autoref{modelstab}. We also use a reference $\Lambda$CDM simulation with the same setup. All simulations share the base cosmological parameters given in \autoref{cosmoparmtab}. Furthermore, they all share the same initial seeds allowing us to divide-out cosmic variance by taking ratios of power spectra.\footnote{ Due to the nonlinear nature of the evolution, some variance could still remain, which could affect these ratios and could artificially improve or degrade the measured accuracy of our predictions. This has been checked by \cite{Cataneo:2018cic} for the evolving DE case using different initial seeds, which found the results to be insensitive to the change in realisation. We have confirmed these results in the dark scattering case, by running smaller simulations (500 Mpc$/h$, 256$^3$ particles) with 2 different random seeds for both $\Lambda$CDM and dark scattering. This exercise has confirmed that the variance in their ratios is always below 1\%, which allows us to safely neglect it.} We have power spectrum measurements up to $k=12~h/{\rm Mpc}$ for $z=0$, $k=9~h/{\rm Mpc}$ for $z=0.5$ and $k=6~h/{\rm Mpc}$ for $z=1$. We refer the interested reader to \cite{Baldi:2016zom} for a more extended description of the simulations and of the modified N-body code.

The 3 different interacting models we consider are only a sample of all the different sets of parameters and time-dependencies for $w$ that are possible in this theory. However, they cover different interaction strengths and values of $w$, two of which ($w$CDM+ and CPL) are relevant in the context of the $\sigma_8$ tension as they reduce its value by approximately $5\%$, matching the discrepancy between CMB and weak lensing data~\citep{Abbott:2021bzy}. Additionally, given that the reaction formalism has been shown by \cite{Cataneo:2018cic} to be accurate for many different functions $w(z)$ for the non-interacting case, we expect that the worsening of accuracy will be a function of the interaction strength, effectively given by $(1+w)\xi$. For this reason, analysing different values of that combination is expected to be relevant for general settings. In spite of this, further cases with different time-dependence would be useful to better understand how the accuracy of our predictions varies with parameter choices. This requires additional simulations, which 
we leave for future work.

We compute our prediction for the matter power using the halo model reaction formalism described above, calculating the reaction $\mathcal{R}$ with the new version of {\tt ReACT} in which the Dark Scattering model is implemented. As the N-body simulation contains only interacting particles, we use the uncorrected coupling, $\xi$, instead of using the effective coupling, $\bar\xi$, given in \autoref{eff_xi} (or equivalently set $f_c=1$).\footnote{ It would be interesting to test the effective coupling approximation in detail, using simulations with more than one type of particle, similar to the ones produced in the recent work by \cite{2022arXiv220104528F}. We plan to test this more realistic situation in future work.} To obtain the full power spectrum, we require an accurate pseudo power spectrum -- a fully nonlinear spectrum in a $\Lambda$CDM cosmology, whose corresponding linear spectrum is equal to that of the cosmology with interaction, at the requested redshift. We obtain that pseudo spectrum using the {\tt EuclidEmulator2} \citep{Knabenhans:2020gdo} by giving as input a scalar amplitude, $\mathcal{A}_s$, corrected by the appropriate growth factors, i.e. $D^2_{\rm DS}(z)/D^2_{\Lambda\rm CDM}(z)$, thus ensuring the match to the linear power spectrum of the dark scattering case. This is possible because this interaction model generates scale independent linear growth, such that a re-scaling of the amplitude is sufficient for the matching. In more complex cases, other alternatives for the pseudo power spectrum may be required, such as {\tt HMCode}, that takes as input the linear power spectrum. We have also tested that option for this work, finding very similar results, but finding the {\tt EuclidEmulator2} to be slightly more accurate on intermediate scales. We present those results in the next section.
\begin{table}
\centering
\caption{A summary of the models considered in this work.}
\begin{tabular}{| c | c | c | c | c | }
\hline  
 Model & $w_0$ &  $w_{a}$  & $\xi \, \left[\frac{\rm {b}}{\rm GeV}\right]$   & $\sigma_8(z=0)$  \\ \hline
 $\Lambda$CDM & -1.0  & 0.0 & 0 & 0.8261  \\ 
 $w$CDM$+$  & -0.9 & 0.0 & 10 & 0.7939 \\ 
 $w$CDM$-$ & -1.1 & 0.0 & 10 & 0.8512  \\ 
 CPL  & -1.1 & 0.3 & 50 &  0.7898 \\ 
 \end{tabular}
\label{modelstab}
\end{table}
\begin{table}
\centering
\caption{Base cosmological parameters.}
\begin{tabular}{| c | c | } 
\hline  
 Parameter & Value \\ \hline
 $h$ &  $0.678$    \\ 
 $\Omega_{\rm c}$ &   $0.2598$  \\ 
 $\Omega_{\rm b}$ &  $0.0482$   \\ 
 $\mathcal{A}_s$ & $2.115 \times 10^{-9}$   \\ 
 $n_s$ & $0.966$ \\ 
 \end{tabular}
 \label{cosmoparmtab}
\end{table}
%
\begin{table}
\centering
\caption{Maximum value of $k$ (in $h/{\rm Mpc}$) for which the residual is below 1\% ($k^{1\%}_{\rm max}$) and 3\% ($k^{3\%}_{\rm max}$).}
\begin{tabular}{| c | c | c | c | c | }
\hline  
  & \multicolumn{2}{c}{$z=0$} & \multicolumn{2}{c}{$z=1$} \\ 
 Model & $k^{1\%}_{\rm max}$ & $k^{3\%}_{\rm max}$ & $k^{1\%}_{\rm max}$ & $k^{3\%}_{\rm max}$ \\ \hline
 $w$CDM$+$ & 1.0 & 1.6 & 3.5 & >6   \\ 
 $w$CDM$-$  & 0.9 & 1.5 & >6 & >6  \\ 
 CPL & 0.8 & 2.0 & 2.0 & 2.6  \\ 
\end{tabular}
\label{resulttab_kmax}
\end{table}

\subsection{Results}\label{sec:wCDM}

We begin by presenting our results for the cases of constant DE equation of state in \autoref{fig:wCDM}, where we show the ratio between the power spectra for the Dark Scattering model and the corresponding $\Lambda$CDM model at three redshifts. We show both the predictions from the reaction formalism as well as from the pseudo power spectrum, in comparison with simulation measurements, for cosmologies with $w = -0.9$ and $w=-1.1$ and a coupling parameter value of $\xi=10$ b/GeV. These two cases have opposite effects, since the interaction parameter, $A$, changes sign depending on whether the value of $\wDE$ lies above or below $-1$, as can be inferred from \autoref{Eq:interaction}. As a result, we see almost symmetric effects extending into the nonlinear scales. There is first an extra suppression (enhancement) at intermediate scales for $w>-1$ ($w<-1$), followed on smaller scales by a much stronger and opposite effect. The first nonlinear regime is very well described by the halo model reaction, with errors below 1\% on those scales. This success is extended to smaller scales for the higher redshifts, but not substantially for $z=0$, when the nonlinear effects of the interaction become larger. In spite of this, at that redshift, our modelling is 1\% accurate for scales up to $k\approx1~h/{\rm Mpc}$ and 3\% accurate up to $k\approx1.5~h/{\rm Mpc}$ for both values of $\wDE$. The detailed values are shown in \autoref{resulttab_kmax}, where the reach of our modelling can also be seen for $z=1$. For that higher redshift our prediction is within 3\% accuracy throughout the range of scales accessible in our simulations, being even within 1\% for the case of $w=-1.1$. This suggests that our modelling can be trusted beyond the scales tested here, but without access to additional simulations (going above $k=6~h/{\rm Mpc}$) we cannot determine a value of $k_{\rm max}$.

In this first case, with $\xi=10$ b/GeV, while the linear effects are considerable, the nonlinear effects of the interaction are fairly small within the scales that we model accurately. This is particularly true at $z=0$, where the size of the effect is sub-percent at $k<1~h/{\rm Mpc}$. For this reason, we have tested whether neglecting the interaction in the reaction calculation would give equally good predictions on these scales. We find that, even for $z=0$, that is not the case and we do need the full model. The reason for this is that in this formulation, nonlinear effects arise both from the pseudo spectrum and from the reaction. With the large linear effects, the pseudo spectrum shows enhanced nonlinear effects on intermediate scales, which are then compensated by the reaction, as seen in \autoref{fig:wCDM}. For that reason, ignoring the interaction in the calculation of the reaction typically doubles the errors relative to simulations around $k\sim1~h/{\rm Mpc}$. Therefore, even when the interaction produces nonlinear effects that are small, their modelling is only accurate when the coupling is fully taken into account.

Next, we validate the final case in \autoref{modelstab}, with a varying equation of state, following the CPL parametrisation. We show our results for that case with $w_0 = -1.1,\, w_a = 0.3$ and $\xi=50$ b/GeV at $z = 0,\, 0.5,\, 1$ in \autoref{fig:CPL}. This case is interesting because the effective coupling, $A$, changes sign at $z=0.5$, first being positive and suppressing linear growth at high redshift, and later enhancing it as redshift goes to 0. As most of the linear clustering occurs before dark energy begins to dominate, the suppression of growth is the largest effect, and the overall result is more similar to the one for $w=-0.9$ than $w=-1.1$. The biggest difference comes in the form of a change of shape on the smallest scales, visible in the results at $z=0$, at which the interaction dampened most of the previous nonlinear amplification. All of these nonlinear effects are enhanced here because we study the much larger interaction strength of $\xi=50$ b/GeV. In spite of this, the prediction of the halo model reaction is 1\% accurate up to scales of $k\approx 0.8~h/{\rm Mpc}$ at $z=0$, reaching $k\approx 2~h/{\rm Mpc}$ at higher redshift, as detailed in \autoref{resulttab_kmax}. In addition, at $z=0$, the errors are never larger than 4\% for the entire range of scales available from the simulations. While this is likely due to the compensation between the effects of the interaction from high and low redshift, it shows that our results capture the correct qualitative behaviour in all cases. In addition, contrary to what happened with the cases with $\xi=10$, here the interaction is responsible for larger nonlinear contributions at all redshifts already on intermediate scales, showing that our modelling is robust in this case too.

In summary, the fact that our predictions are accurate for a substantial range of scales and redshifts, particularly for large interaction strengths, demonstrates that the reaction formalism is effective at modelling the nonlinear effects of interacting dark energy and can be used for the analysis of real data to constrain the interaction strength. For that to be fully realised, however, we need to understand all the contributions to the power spectrum on small scales, including those generated by baryon feedback and massive neutrinos. We analyse that in the next section.

\section{Including baryonic feedback and massive neutrinos}\label{sec:bar_neu}

In the previous section, we have shown that we can model the effects of interacting dark energy on nonlinear scales with percent-level accuracy, as show in \autoref{resulttab_kmax}. However, to get a full prediction for the matter power spectrum that is relevant for observations, we need to include the effects of baryon feedback and massive neutrinos, which substantially alter the power spectrum on small scales.

Both these effects represent a scale-dependent suppression of power that could imitate the features of interacting dark energy. As seen in \autoref{fig:wCDM} and \autoref{fig:CPL}, the nonlinear effect of the interaction initially enhances the linear effect at intermediate scales, representing a suppression for $w>-1$ and an enhancement for $w<-1$. On highly nonlinear scales, however, the effect is reversed, as the additional friction for $w>-1$ causes structures to lose energy and collapse to deeper potential wells, thus forming denser structures, with the inverse happening for $w<-1$. This is similar to what occurs with baryonic feedback, which induces suppression of power on intermediate scales, due to gas expulsion from AGN feedback, and an enhancement on smaller scales due to star formation. It is therefore likely that there is a degeneracy between the two effects. Massive neutrinos also induce a suppression of power, since they do not cluster as efficiently as cold matter on sufficiently small scales. While this effect is less similar to that of the interaction, it could also be somewhat degenerate, particularly on intermediate scales.

In this section we describe how these effects can be modelled within the reaction formalism, and perform a first analysis of the degeneracy between them and the contributions of the interaction. Throughout this section, we employ the effective coupling approximation of \autoref{eff_xi}, as there would be no realistic scenario for which only dark matter is present.

\subsection{Baryonic feedback}\label{sec:baryons}

Following \cite{2021MNRAS.508.2479B}, we include the effects of baryonic feedback in the computation of the pseudo power spectrum via {\tt HMCode} \citep{Mead:2020vgs}.\footnote{This could alternatively be done by using promising emulators of baryonic effects, {\tt BCMemu} \citep{2021arXiv210808863G}, and {\tt bacco} \citep{Arico:2020lhq}.} The model implemented there is also based on the halo model and takes into account effects from Active Galactic Nuclei (AGN) feedback and star formation, encoding them in a 6-parameter model, accounting for time evolution. That model was then fitted to the {\tt BAHAMAS} simulations \citep{McCarthy:2016mry} to obtain a 1-parameter model, which depends only on the temperature of AGN, via $\theta \equiv \log_{10} (T_{\rm AGN}/\rm K)$, and was validated in the range $7.6 \leq \theta \leq 8.3$ \citep[see][for more details]{Mead:2020vgs}. This model is called {\tt HMCode2020$\_$feedback} and is the one we use in this work.
\begin{figure}
    \centering
    \includegraphics[width=0.475\textwidth]{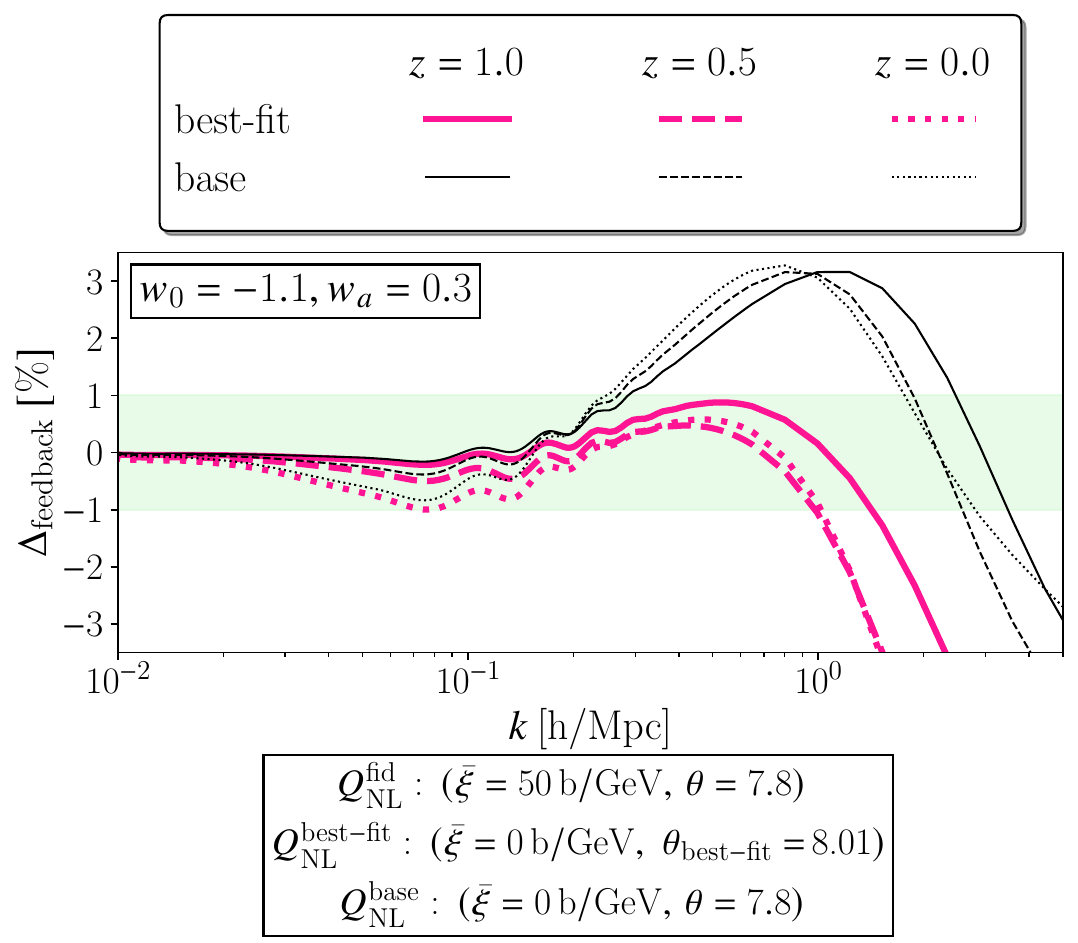}
    \caption{Residuals of the comparison between the Dark Scattering predictions for $\bar\xi=50$ b/GeV ($\xi=60$ b/GeV), $\theta = 7.8$ and two non-interacting cases with different baryonic feedback, at three different redshifts $z = 1$ (solid), $z = 0.5$ (dashed), and $z = 0$ (dotted). In pink, we show the best-fit case ($\theta_{\rm best-fit} = 8.01$) and in black the base case, with the same temperature as the interacting case. In all cases we use the same parameters as the CPL case: $w_0=-1.1$, $w_a = 0.3$. The residual is defined by $\Delta_{\rm feedback} = 100\,\mathcal{\%} \cdot \left( 1 - Q^{\rm fid}_{\rm NL}/Q_{\rm NL}^{\rm non-int}\right)$, with non-int being best-fit or base depending on the case.}
    \label{fig:my_label5}
\end{figure}
The correction for baryonic effects is taken into account by an extra boost factor, $\mathcal{B}(k,z)$, multiplying the power spectrum, defined by the ratio between the full power spectrum and the DM-only spectrum, as follows,
\bea
\mathcal{B}(k,z) = \dfrac{P_{\rm full}}{P_{\rm DM\text{-}only}} \, .
\eea
Therefore, to obtain the final power spectrum within the halo model reaction framework, we combine {\tt HMCode2020$\_$feedback} and {\tt ReACT} prescriptions to obtain,\footnote{This is the general prescription for obtaining the power spectrum when the three ingredients are generated by different methods. Naturally, when the full pseudo power spectrum is created also by {\tt HMCode}, there is no need to separate the boost out.}
\bea
P_{\rm NL}(k,z) =   \mathcal{R}(k,z) \times \mathcal{B}(k,z) \times P^{\rm pseudo}_{\rm DM\text{-}only} (k,z) \, .
\label{full:spectrum}
\eea
With this prescription we can readily predict the nonlinear matter power spectrum for Dark Scattering, in the presence of baryon feedback effects.\footnote{Note that, while we do not test the validity of using a $\Lambda$CDM prescription for the baryon boost, this has been tested by \cite{2021MNRAS.508.2479B} in the non-interacting case and found to be an accurate prescription. Additionally, it has been shown that the baryonic boost has nearly no dependence on cosmology, except for the baryon fraction~\citep{McCarthy:2017csu,vanDaalen:2019pst,2021arXiv210808863G}, and, given that the interaction does not affect baryons, we expect the same to apply in this case.}

We now focus in determining whether a degeneracy exists between IDE and baryonic feedback parameters at the level of the spectrum in \autoref{full:spectrum}. While it is clear that the linear effects of the interaction do not have a counterpart baryonic effect, they are degenerate with the amplitude of the power spectrum as well as with $w$, given their scale-independence. At nonlinear scales, however, the contributions from baryons and IDE could be very similar or partially cancel each other out, so this degeneracy needs to be understood. To that end, we will attempt to fit a non-interacting model with varying baryon feedback to an interacting model with fixed baryon feedback, in order to ascertain whether the baryonic effects can mimic the interaction contribution. To isolate the scale-dependent nonlinear effects we are interested in, instead of comparing the full power spectra, we match instead spectra normalized to their large scale value, 
\bea
Q_{\rm NL}\equiv \frac{P_{\rm NL}}{P_{\rm NL}(k_*)}\,,
\eea
where $k_*\ll k_{\rm NL}$ so that $P_{\rm NL}(k_*)\approx P_{\rm L}(k_*)$.  We begin by generating our fiducial spectrum with $\bar\xi = 50\, \rm {b} /\rm GeV$ ($\xi=60$ b/GeV), $w_0 = -1.1$, $w_a=0.3$,\footnote{We chose the case with the largest interaction strength we had validated since it is the case generating the largest nonlinear effects and therefore the one in which the degeneracy will be most clearly seen. In addition, as described in \autoref{sec:wCDM}, the nonlinear effects of $\xi = 10\, \rm {b} /\rm GeV$ models are small and even if they could be fully mimicked by the baryonic effects, a degeneracy could not be claimed, as it would be within the expected errors in the theoretical modelling.} adding a baryon boost with $\theta = 7.8$ and thus computing $Q_{\rm NL}^{\rm fid}$. We then generate results without interaction ($\xi = 0$), while varying feedback strength until we are able to find a value $\theta_{\rm best-fit}$ which minimises the residuals between this case and the fiducial spectra. We fit all three redshift bins together and we consider only the scales for which our predictions from the halo model reaction have an accuracy of $\Delta \leq 1 \%$ (displayed in \autoref{fig:CPL}), ensuring that the feedback is not mimicking incorrect effects. As the scale-dependent effect of the interaction in the chosen case is to suppress power, we expect the baryon feedback model that best mimics it to have a higher temperature than the fiducial case. The fiducial temperature of $\theta=7.8$ was chosen taking this fact into account, both ensuring that itself and the higher value resulting from the fitting procedure are within the well-tested region of parameter space ($7.6 \leq \theta \leq 8.3$).

The results from this procedure are illustrated in \autoref{fig:my_label5}, where we show the residuals between the fiducial and best fit cases, as well as compared to a base non-interacting case with the same temperature as the interacting case. As seen there the best-fit value is $\theta_{\rm best-fit} = 8.01$, which is substantially different from the fiducial value of 7.8. It is also clear in the figure that modulating the feedback strength can mimic the effect of the interaction, as it reduced the residuals to below 1\% on scales up to $k=1~h/{\rm Mpc}$, in which we trust our modelling fully, whereas the original residual is above 3\% on the same scales, at all redshifts, as seen in the black lines. This clearly suggests that the two effects are degenerate and that they would be hard to distinguish in an analysis including only these scales. Our modelling seems to indicate that going to smaller scales is likely to dilute this degeneracy, but that is also where the errors in our predictions begin to grow, indicating that better modelling may be required to completely understand this degeneracy. The presence of this degeneracy highlights the importance of using a combined analysis with spectroscopic clustering to constrain interacting dark energy (as analysed by \cite{Carrilho:2021hly}), as their relative independence of baryonic feedback would help in breaking this degeneracy.

\subsection{Massive neutrinos}\label{sec:neutrinos}
We now aim to investigate the degeneracy between massive neutrinos and the dark sector interaction. The effect of massive neutrinos is fully included in the reaction formalism, as described in \autoref{sec:reaction} and it is therefore straightforward to compute using {\tt ReACT}. We proceed in the same way as in the previous section, attempting to find a value for the neutrino mass for which $Q_{\rm NL}$ mimics that of the interacting cosmology. Note that, while for the baryon feedback case, all the effects measured by $Q_{\rm NL}$ were purely nonlinear, in the case of neutrinos, $Q_{\rm NL}$ also includes scale-dependent linear effects, which are not degenerate with the scalar amplitude, and therefore have to be included for a consistent study of degeneracy with the nonlinear effects of the interaction.

\begin{figure}
    \centering
    \includegraphics[width=0.475\textwidth]{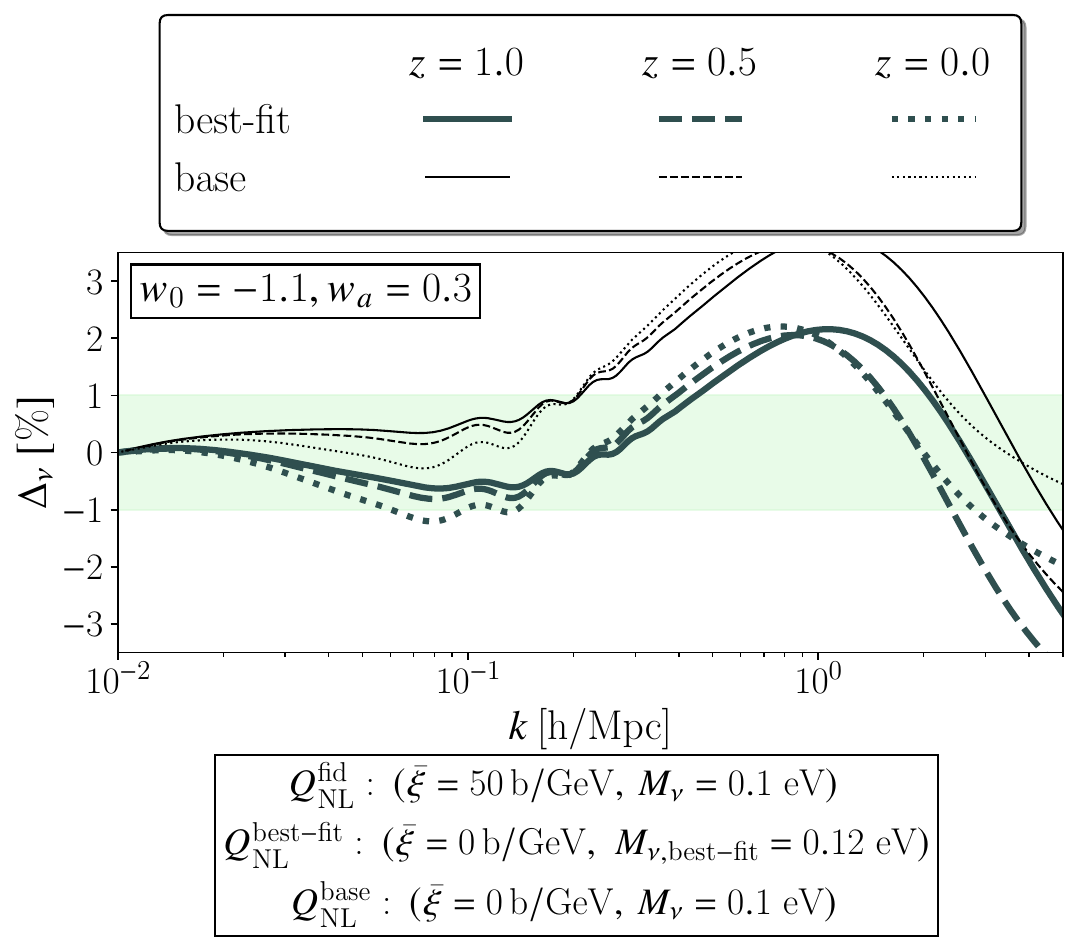}
    \caption{Residuals of the comparison between the Dark Scattering predictions for $\bar\xi=50$ b/GeV ($\xi=60$ b/GeV), $M_\nu = 0.1\,$ eV and two non-interacting cases with different neutrino mass, at three different redshifts $z = 1$ (solid), $z = 0.5$ (dashed), and $z = 0$ (dotted). In dark green, we show the best-fit case ($M_{\nu,{\rm best-fit}} = 0.12~{\rm eV}$) and in black the base case, with the same neutrino mass as the interacting case. In all cases we use the same parameters as the CPL case: $w_0=-1.1$, $w_a = 0.3$. The residual is defined by $\Delta_\nu = 100\,\mathcal{\%} \cdot \left( 1 - Q^{\rm fid}_{\rm NL}/Q_{\rm NL}^{\rm non-int}\right)$, with non-int being best-fit or base depending on the case.}
    \label{fig:massive_nu}
\end{figure}

In \autoref{fig:massive_nu}, we show the percentage deviation from a fiducial spectrum with $\bar\xi = 50\, \rm {b} /\rm GeV$ ($\xi=60$ b/GeV), $w_0 = -1.1$, $w_a=0.3$ and a neutrino mass of $M_\nu = 0.1$ eV to a no-interaction spectrum having $\xi =~0~\rm {b} /\rm GeV$, $w_0 = -1.1$, $w_a=0.3$ but a neutrino mass of $M_\nu = 0.12$~eV. We also show the comparison to a base non-interacting case with the same neutrino mass as the interacting case.  We find over the three redshifts considered, this neutrino mass is the one that fits the fiducial best in the absence of interaction in the dark sector. We clearly see that a dark sector momentum exchange at the level of $\bar\xi = 50 \, \rm {b} /\rm GeV $ cannot be mimicked by massive neutrinos alone, even if a single redshift slice is considered in isolation. This is in contrast with the baryonic feedback case where we can achieve better agreement even when considering the entire redshift range.\\

It should be noted that we use a single set of dark energy parameters to test both degeneracies, and it is conceivable that a very different dark energy evolution would change these conclusions. However, the generic feature that the interaction creates on small scales scales is expected to be fairly robust, given that it is observed in all our results. Still, more detailed forecasts of these degeneracies are necessary also taking into account the detection capabilities of future surveys. These will have to include an analysis of how this degeneracy depends on dark energy evolution and we leave them for future work.

%
\section{Summary}\label{sec:summary}

In this paper we have used the halo model reaction formalism to build a complete model for the nonlinear matter power spectrum for the dark scattering model. We did this by extending the halo model to include the additional force acting on dark matter particles, which required modifying the spherical collapse dynamics, as well as the virial theorem, in addition to the linear evolution. These extensions have been implemented into {\tt ReACT} \citep{Bose:2020wch}, which is publicly available (see \hyperref[data_av]{Data Availability}).

To our knowledge, this represents the first analytical model for the nonlinear power spectrum for an interacting dark energy model. We conjecture that many other models of momentum-exchange interactions could be readily described by this formalism, simply by modifying the time-dependence of the interaction function, $A$, in \autoref{Eq:interaction}. Additionally, much of the formalism constructed here would be useful to generate nonlinear predictions for more complex interacting models. Those extensions will be explored in future work.

We have validated our modelling against simulations with the dark scattering interaction from \cite{Baldi:2016zom}, using a pseudo spectrum generated with the {\tt EuclidEmulator2} \citep{Knabenhans:2020gdo}. We found that our predictions have 1\% agreement with simulations for scales up to $O(1)~h/{\rm Mpc}$ at redshift zero, improving beyond that at higher redshift, as summarized in \autoref{resulttab_kmax}. In particular, at $z=1$, the accuracy is only worse than 1\% after $k=2~h/{\rm Mpc}$. This includes both models with very mild nonlinear effects of the interaction ($\xi=10$ b/GeV), but also a case for which they are much larger ($\xi=50$ b/GeV). This allows us to conclude that our modelling is successful at describing the nonlinear power spectrum and can be used for analysing data. 

In order to further extend the reach into even smaller scales, additional steps could be taken. This includes using a more accurate concentration-mass relation, fitted to simulations, which was shown by \cite{Cataneo:2019fjp} to improve the accuracy. In addition, improving the modelling of angular momentum loss during collapse could also enhance the accuracy, given that this is a crucial contributor to the effects of dark scattering on the smallest scales \citep{Baldi:2014}.

We have also included the effects of baryonic feedback using {\tt HMCode} \citep{Mead:2020vgs}; as well as massive neutrinos, using the full reaction formalism \citep{2021MNRAS.508.2479B}. We can thus generate predictions for the full power spectrum that is directly probed by experiment. We have analysed the degeneracies between the interaction and those two ingredients by attempting to mimic the nonlinear effects of dark scattering by varying the baryon feedback parameter or the neutrino mass. We find that limiting the scales to those for which our accuracy is within 1\% reveals a degeneracy in the case of baryon feedback. Extending beyond those scales, however, we expect that the degeneracy could be broken, given the stronger nonlinear effects generated in the interacting model. For the case of massive neutrinos, no significant degeneracy is found.  While these simplified tests already reveal some potential degeneracies, a more thorough MCMC analysis would enable us to exactly pin them down, as well as allowing us to fully validate our nonlinear modelling. We aim to fully explore that in future work, where we will also be able to precisely define the range of applicability of our modeling as well as forecast the observability of the dark scattering effects shown here with photometric surveys.

\section*{Acknowledgments}
\noindent We are grateful to Matteo Cataneo and Catherine Heymans for feedback. P.C.'s research is supported by a UK Research and Innovation Future Leaders Fellowship. A.P. is a UK Research and Innovation Future Leaders Fellow [grant number MR/S016066/1]. A.P. is grateful to the Cosmic Dawn Center, University of Copenhagen, for hospitality while part of this work was in progress. K.C. acknowledges support from a CONACyT studentship. J.C.H. and K.C. acknowledge support from program UNAM-PAPIIT [grant number IN-107521]. B.B. and L.L. acknowledge support from the Swiss National Science Foundation (SNSF) Professorship grant No.~170547. M.B. acknowledges support by the project “Combining cosmic microwave background and large scale structure data: an Integrated approach for addressing fundamental questions in cosmology" funded by the PRIN-MIUR 2017 grant 2017YJYZAH. 
We acknowledge the use of open source software \citep{scipy:2001, Hunter:2007, mckinney-proc-scipy-2010, numpy:2011}.
\appendix 


\section*{Data Availability}
\phantomsection
\label{data_av}

The data underlying this article will be shared on reasonable request to the corresponding author. 
The software employed is publicly available in {\tt ReACT} \github{https://github.com/PedroCarrilho/ReACT/tree/react_with_interact_baryons}. The DM-only pseudo power spectrum is provided by the {\tt EuclidEmulator2} \github{https://github.com/miknab/EuclidEmulator2}.The baryonic feedback correction is provided by {\tt HMCode} \github{https://github.com/alexander-mead/HMcode}. The linear predictions are obtained by a modified version of {\tt CLASS} \github{https://github.com/PedroCarrilho/class_public/tree/IDE_DS}.


\bibliographystyle{mnras}
\bibliography{mybib} 

\bsp	
\label{lastpage}
\end{document}